\documentclass[twocolumn,prd,amsfonts,showpacs,epsfig,floats,superscriptaddress,floatfix]{revtex4}
\usepackage{graphicx}
\usepackage{dcolumn}
\usepackage{bm}
\usepackage{longtable}
\usepackage{amsmath}
\usepackage{mathrsfs}

\usepackage{epstopdf}
\graphicspath{{./}}

\begin{document}
\bibliographystyle{apsrev}


\title{Affine equation of state from quintessence and k-essence fields}

\author{Claudia Quercellini}
\affiliation{Dipartimento di Fisica, Universit\`a di Roma ``Tor Vergata'',
via della Ricerca Scientifica 1, 00133 Roma, Italy}
\author{Marco Bruni}
\affiliation{Dipartimento di Fisica, Universit\`a di Roma ``Tor Vergata'',
via della Ricerca Scientifica 1, 00133 Roma, Italy}
\affiliation{INFN Sezione di Roma ``Tor Vergata'',
via della Ricerca Scientifica 1, 00133 Roma, Italy}
\affiliation{Institute of Cosmology and Gravitation, University
of Portsmouth, Mercantile House, Portsmouth PO1 2EG, Britain}
\author{Amedeo Balbi}
\affiliation{Dipartimento di Fisica, Universit\`a di Roma ``Tor Vergata'',
via della Ricerca Scientifica 1, 00133 Roma, Italy}
\affiliation{INFN Sezione di Roma ``Tor Vergata'',
via della Ricerca Scientifica 1, 00133 Roma, Italy}

\begin{abstract}
We explore the possibility that a scalar field with appropriate Lagrangian can mimic a perfect fluid with an affine barotropic equation of state. The latter can be thought of as a generic cosmological dark component evolving as an effective cosmological constant plus a generalized dark matter. As such, it can be used as a simple, phenomenological model for either dark energy or unified dark matter. Furthermore, it can approximate (up to first order in the energy density) any barotropic dark fluid with arbitrary equation of state. We find that two kinds of Lagrangian for the scalar field can reproduce the desired behaviour: a quintessence-like with a hyperbolic potential, or a purely kinetic k-essence one. We discuss the behaviour of these two classes of models from the point of view of the cosmological background, and we give some hints on their possible clustering properties.
\end{abstract}

\pacs{98.80.-k; 98.80.Jk; 95.35.+d; 95.36.+x}
\maketitle

\section{Introduction}\label{intro}

Known forms of matter such as baryons and radiation are not sufficient to explain the observed universe, hence we need to assume the existence of an unknown dark component, whose properties are then deduced from indirect detections. About one third of it, named dark matter, is needed to account for the inhomogeneities that we observe up to very large scales, namely to explain both large scale structure and the cosmic microwave background anisotropy peaks \cite{Spergel:astro-ph/0603449,2006Natur.440.1137S}. The remaining two thirds, dubbed dark energy, are needed to explain the observed flatness of the universe and its late time acceleration \cite{Perlmutter:astro-ph/9812133,Riess:astro-ph/0611572}. Assuming the existence of heavy particles, non collisional and cold,  dark matter is usually modelled as a pressureless perfect fluid.  In its simplest form dark energy  takes the form of  vacuum energy density, i.e.\ a cosmological constant $\Lambda$. More generally, it can be modeled  as a perfect fluid with an equation of state (EoS from now on) that can violate the strong energy condition (SEC, see e.g.\ \cite{Visser:1997aa}), such  that it can dominate at late times and  have sufficiently negative pressure to account for  the observed accelerated expansion. For both forms of energy several tentatives have been done in different frameworks to make them descend from a scalar field, related to a  Lagrangian  (see \cite{Copeland:hep-th/0603057v3} for a recent review and references therein). 
Since we do not know much about these dark components, the idea that they may be ascribed to  a unique source of energy is appealing, especially if their origin can be easily related to some fundamental theory, as it  is for a scalar field. A considerable effort at describing the dark side of the universe with a unified phenomenological model \cite{Kamenshchik:gr-qc/0103004,Bento:astro-ph/0407239} has been made in the last few years, also aiming at constraining the parameters of the models (e.g. see  \cite{Balbi:astro-ph/0702423,Amendola:astro-ph/0304325}). In this context scalar field models have been proposed, both with  a canonical kinetic term in the  Lagrangian \cite{Sahni:astro-ph/9910097,Mainini:astro-ph/0503036v2} and in k-essence scenarios \cite{Giannakis:astro-ph/0501423v2, Beca:astro-ph/,Chimento:astro-ph/0311613,Scherrer:astro-ph/0402316,bertacca:astro-ph/0703259}.

In this paper we develop a  different approach: we investigate whether a minimally coupled scalar field with a specific  Lagrangian can mimic  the dynamics of the dark fluid discussed in \cite{Balbi:astro-ph/0702423,Visser:gr-qc/0309109,Chiba:astro-ph/9704199,Gorini:hep-th/0311111,Ananda:astro-ph/0512224} and similarly in \cite{babichev:arXiv:astro-ph/0407190,holman:astro-ph/0408102}. More precisely, in \cite{Balbi:astro-ph/0702423} the working hypothesis was the existence of a single dark perfect fluid with a simple 2-parameter barotropic equation of state, the affine EoS $P_{X}=p_{0}+\alpha\rho_{X}$, which naturally yields an energy density evolution of an effective cosmological constant plus a generalized dark matter \cite{Ananda:astro-ph/0512224}. This EoS can also  be seen  as a first order Taylor expansion  of a wider class of functions of the energy density, where the dark component more generally can either only provide the late time acceleration or mimic also a matter-like background evolution. In addition, this EoS can be derived from the simple assumption that the speed of sound is constant and potentially positive, alleviating the unpleasant  consequences of a negative value for structure formation \cite{Hu:astro-ph/9801234}.  In \cite{Balbi:astro-ph/0702423} we tested this model against observables related to the homogeneous and isotropic  expansion and compared it to the standard $\Lambda$CDM model.  Here we will investigate if and to what extent such a EoS, in general representing either a dark energy or a unified dark matter  component,  can arise  from scalar field dynamics, considering two possibilities: a quintessence model, with a hyperbolic potential, and a purely kinetic k-essence model.

\section{Affine equation of state}
\label{affine}
In an homogeneous isotropic universe, modeled with a Robertson-Walker metric, the energy momentum tensor must take the perfect fluid form, with a total energy density $\rho_T = \sum_i \rho_i$ and pressure $P_T=\sum_i P_i$. Under standard assumptions,  if the various components are non-interacting, each satisfies the usual conservation equations, $\nabla_\mu T^{\mu\nu}_{(i)}=0$,
independently of the theory of gravity. 
To fix ideas, let us assume that $X$, one of the above components,
is represented by a barotropic fluid with EoS $P_X=P_X(\rho_X)$. Let's assume that this EoS allows for violation of SEC  at least below some redshift, so that $X$ can become the dominant component and drive the observed cosmic acceleration. Assuming now and in the following Einstein equations with no cosmological constant $\Lambda$, violation of SEC is equivalent to $\rho_X+3P_X\leq 0$, which in a Friedmann-Robertson-Walker universe is sufficient for acceleration.  Then, denoting with  $H$  the Hubble expansion scalar related to the scale factor $a$ by $H=\dot{a}/a$, it follows from   the  energy conservation equation
\begin{equation}
\label{einstein2}
\dot{\rho}_X=-3H(\rho_X+P_X)
\end{equation}
 that, if  there exists  an energy density value  $\rho_X=\rho_\Lambda$ such that $P_X(\rho_\Lambda)=-\rho_\Lambda$, then $\rho_\Lambda$ has the dynamical role of an {\it effective} cosmological constant: $\dot{\rho}_\Lambda=0$ (see \cite{Ananda:astro-ph/0512224} for a more detailed discussion).
 
Another reasonable assumption is that the square of the speed of sound of $X$ is non-negative: $c_X^2:=dP_X/d\rho_X\geq 0$. Indeed, this assumption ensures that the adiabatic perturbations of $X$ do not blow up (because it follows from momentum conservation that, for a barotropic fluid and for $dP/d\rho<0$, pressure gradients do not work anymore as a restoring force against gravity, and instead act {\it as} gravity). Actually, the assumption $c_X^2\geq0$ is strong enough to imply violation of SEC and  a sort of cosmic no-hair theorem, if $P_X<0$ at some point (in an expanding phase, $H>0$, with $\rho_X+P_X>0$).
That is, under these conditions it follows that the energy density decreases while the pressure becomes negative enough for SEC to be violated, thereby driving an accelerated expansion in a Friedmann-Robertson-Walker universe, till $\rho\rightarrow\rho_\Lambda$, so that the universe becomes de Sitter at late times, i.e. $\rho_\Lambda$ is an attractor for (\ref{einstein2}). When $\rho_\Lambda$ becomes the dominant component, the same holds true in Bianchi models, as proved by Wald \cite{Wald:1983ky}, as well as in other cases, see e.g.\ \cite{Wainwright:2005}.

In \cite{Balbi:astro-ph/0702423} we have considered a flat Friedmann-Robertson-Walker universe in general relativity, with radiation, baryons and a single unified dark matter component with energy density $\rho_X$ represented by a barotropic fluid. 
Given that the EoS $P_X=P_X(\rho_X)$ is unknown, we assumed a constant speed of sound  $dP_X/d\rho_X\simeq \alpha$, leading to the  2-parameter affine form \cite{Ananda:astro-ph/0512224}
\begin{equation}
\label{EOSaffine}
P_X\simeq p_0+\alpha \rho_X.
\end{equation}
This allows for violation of SEC even with \mbox{$c_s^2=\alpha \geq 0$}. Then, using (\ref{EOSaffine}) in (\ref{einstein2})  and asking for $\dot{\rho}_\Lambda=0$ leads to the {\it effective cosmological constant } $\rho_\Lambda=-p_0/(1+\alpha)$.
Eq.\ (\ref{EOSaffine}) may also be regarded (after regrouping of terms) as the Taylor expansion, up to ${\cal O}(2)$, of {\it any} EoS $P_X=P_X(\rho_X)$ about the present energy density value $\rho_{Xo}$ \cite{Visser:gr-qc/0309109}.
The EoS (\ref{EOSaffine}), if taken as an approximation, could be used to parametrize a dark component (either unified dark matter or dark energy) at low and intermediate redshift. Actually, with  $\alpha \rightarrow 0$  (and  ${\rho}_m>0$, see Eq.\ (\ref{rhodia}) below) the EoS above  is  equivalent to a $\Lambda$CDM. This allows for a straightforward comparison of models, as done in \cite{Balbi:astro-ph/0702423}. Indeed, in \cite{Balbi:astro-ph/0702423} we made a more radical assumption, that is, we extrapolated the validity of Eq.\ (\ref{EOSaffine}) to any time, thereby building a cosmological model based on a  unified dark  component with EoS (\ref{EOSaffine}), which we tested against observations and compared with the standard $\Lambda$CDM model. We found that $\Omega_\Lambda\simeq 0.7$ and $\alpha\simeq 0.01$.
 
 The evolution of $\rho$ with the expansion can be found 
 using the EoS (\ref{EOSaffine}) in the conservation equation \ (\ref{einstein2}), leading to
\begin{equation}
\label{rhodia}
\rho_X(a)=\rho_\Lambda+\rho_m a^{-3(1+\alpha)},
\end{equation}
where today $\rho_m=\rho_{Xo}-\rho_\Lambda$ and $a=1$. Formally, with the EoS (\ref{EOSaffine}) one can then interpret our  dark component  as made up of the {\it effective} cosmological constant $\rho_\Lambda$ and an evolving part with present ``density" ${\rho}_m$. It is clear that the standard $\Lambda$CDM model is recovered for $\alpha=0$,  if $\rho_m>0$ is identified with the density of pressureless dark matter. More in general, 
{\it a priori} no restriction on the values of $\alpha$ and $p_o$ is required, but  one needs $p_o<0$ and $\alpha>-1$ in order to satisfy the conditions that $\rho_\Lambda>0$ and   $\rho \rightarrow \rho_\Lambda$ in the future, i.e. to have that $\rho_\Lambda$ is an attractor for Eq.\ (\ref{EOSaffine}). In this case, our affine dark matter is phantom 
if ${\rho}_m <0$, but without a ``big rip", cf.\ \cite{Ananda:astro-ph/0512224}.
 
 In summary, the barotropic affine EoS (\ref{EOSaffine}) can be used to model either a dark energy component, if standard dark matter is also assumed to be present, or a unified dark matter, as we did in \cite{Balbi:astro-ph/0702423}. 
 
It is well known that, from a given EoS describing a fluid, a scalar field model can be derived (see e.g. \cite{1991CQGra...8..667E}). However, the correspondence is in general not one to one: scalar fields in general have an extra degree of freedom and thus a more complicated dynamics. The question then arises, whether this dynamics allows for a solution which acts somehow as an attractor, so that it can mimic, at least to a certain extent and with little or no fine tuning, the fluid evolution. 
 In the following sections we are going to investigate to what extent a scalar field with appropriate Lagrangian can mimic the cosmological dynamics arising from the affine EoS 
 (\ref{EOSaffine}), i.e. the density evolution (\ref{rhodia}). We shall assume a flat Friedmann-Robertson-Walker universe with units $c=1$ and $8\pi G=1$, so that the Hubble parameter $H^{2}=\dot{a}/a$ and the energy density $\rho$ obey the Friedmann equation $H^{2}=\rho / 3$.


\section{Scalar fields}
\label{SF}
In the most general form, in general relativity, the action for a minimally coupled scalar field can be written as
\begin{equation}
\label{action}
S=\int d^{4}x \sqrt{-g}\Big(\frac{R}{2}+\mathcal{L}(\chi,\phi)\Big),
\end{equation} 
where $\chi=-\frac{1}{2}g^{\mu\nu}\partial_{\mu}\phi\partial_{\nu}\phi$ is the kinetic term. The stress energy tensor  consequently is 
\begin{equation}
\label{tensor1}
T_{\mu\nu}^{\phi}=\frac{\partial \mathcal{L}(\phi,\chi)}{\partial \chi}\partial_{\mu}\phi \partial_{\nu}\phi+\mathcal{L}(\phi,\chi)g_{\mu\nu}
\end{equation} 
and, as long as the scalar field 4-gradient is time-like, i.e.\ $g^{\mu\nu}\partial_{\mu}\phi \partial_{\nu}\phi<0$, it can be cast in the perfect fluid form $T_{\mu\nu}^{\phi}=(\rho_{\phi}+p_{\phi})u_{\mu}u_{\nu}+p_{\phi}g_{\mu\nu}$, where $u_{\mu}=\frac{\partial_{\mu}\phi}{\sqrt{2\chi}}$ is the 4-velocity (coinciding with the unit normal to the $\phi=constant$ slices) and the  Lagrangian plays the role of the pressure of the fluid. For an observer comoving with the fluid, the energy density  reads
\begin{equation}
\label{density}
\rho_{\phi}(\phi,\chi)=2\chi\frac{\partial p_{\phi}(\phi,\chi)}{\partial \chi}-p_{\phi}(\phi,\chi),
\end{equation} 
while the effective EoS and the speed of sound  \cite{Garriga:hep-th/9904176} are
\begin{equation}
\label{cs2}
w_{\phi}=\frac{p_{\phi}(\phi,\chi)}{2\chi\frac{\partial p_{\phi}(\phi,\chi)}{\partial \chi}-p_{\phi}(\phi,\chi)};\qquad c_{\phi}^{2}=\frac{\partial p_{\phi}(\phi,\chi)/\partial \chi}{\partial \rho_{\phi}(\phi,\chi)/\partial \chi}.
\end{equation} 
Thus a scalar field admits a perfect fluid description, in general {\it not} of barotropic type. Indeed, comparing with a barotropic fluid, a scalar field has an  extra degree  of freedom: therefore, in general  each initial condition (equivalently, each trajectory in phase space) for a given scalar field Lagrangian corresponds to a different barotropic EoS. Hence, the question is to what  extent a given EoS can be mimicked by a given scalar field model,  at least in a asymptotic regime, i.e.\ to what extent that model admits a solution that acts as attractor for other trajectories in phase space, thereby avoiding a strong fine tuning problem that would spoil the value of the model itself.

 In the next subsections we will consider separately two scalar field models: we will first analyse a quintessence model and derive a potential that returns the affine EoS as an exact solution in phase space, then we will reconstruct the pressure in a purely k-essence model where the potential is set to be constant.

\subsection{Quintessence}
\label{sec_q}
 In the quintessence model  the  Lagrangian is just the difference between a canonical kinetic term and a potential,
 \begin{eqnarray}\label{lag_q}
\mathcal{L}=\chi-V(\phi)=\lambda\frac{1}{2}\dot{\phi}^{2}-V(\phi)
\end{eqnarray}
at the zero order, and the energy density and pressure take the form
 \begin{eqnarray}\label{den_q}
p_{\phi}=\mathcal{L}\qquad \rho_{\phi}=\lambda\frac{1}{2}\dot{\phi}^{2}+V(\phi),
\end{eqnarray}
where $\lambda=1$ for the standard quintessence scenario and $\lambda=-1$  for phantom cases (not yet excluded by current observations \cite{Seljak:astro-ph/0604335}). The Klein Gordon equation for the scalar field is $\ddot{\phi}+3H\dot{\phi}+V'(\phi)=0$, which will be analysed later in this section. Combining Eq.~(\ref{den_q}) with  Eq.~(\ref{EOSaffine})  we can express the derivative of the field and the potential energy as functions of the scale factor:
\begin{eqnarray}\label{eq:system1}
\lambda\dot{\phi}^2&=& {\rho}_m(1+\alpha)a^{-3(1+\alpha)}\\
\label{eq:system2}
V(\phi)&=& {\rho}_\Lambda+ \frac{{\rho}_m(1-\alpha)}{2}a^{-3(1+\alpha)}.
\end{eqnarray} 
The dot here refers to the derivative with respect to cosmic time $t$. Phantom evolution for the energy density, namely solutions for which the energy density is a growing function of the scale factor, can be set by either $\alpha<-1$ or $\rho_{m}<0$.

We now want to work out the shape of the potential integrating and inverting Eq.~(\ref{eq:system1}), eventually substituting it in Eq.~(\ref{eq:system2}).
Let us assume that  the fluid is the only dominating component (i.e. $H^2=\rho_\phi/3$) and rewrite  Eq.~(\ref{eq:system1}) using $\frac{d\phi}{dt}=aH\frac{d\phi}{da}$:
\begin{equation}\label{eq:dpda}
\frac{d\phi}{da}=\sqrt{\frac{3(1+\alpha)}{\lambda a^{2}}\frac{1}{(1+\frac{\rho_{\Lambda}}{\rho_{m}}a^{3(1+\alpha)})}},
\end{equation}
where we choose the positive sign for the square root of $H^{2}$, connected to a universe that is expanding at least at present time, although one can also have a change of sign in this function (see point {\em iii)} below). Also, the outcome of our calculations will be independent of the sign of $\frac{d\phi}{da}$, thanks to the simmetry of the potential about its minimum (see below).
 Integrating out Eq.~(\ref{eq:dpda}) we find
\begin{equation}\label{eq:pda}
\sqrt{\frac{3(1+\alpha)}{\lambda}}\phi=-2\log{\Big(a^{-\frac{3(1+\alpha)}{2}}+\sqrt{\frac{\rho_{\Lambda}}{\rho_{m}}+a^{-3(1+\alpha)}}\Big)},
\end{equation}
which inverted and substituted in Eq.~(\ref{eq:system2}) returns the expression for the potential:
\begin{eqnarray}\label{eq:vdip}
V(\phi)&=& \Big[\frac{3+\alpha}{4}\rho_{\Lambda}+\frac{(1-\alpha)}{8}(\rho_{m}e^{-\sqrt{3\lambda(1+\alpha)}\phi}\\
&+&\Big(\frac{\rho_{\Lambda}^{2}}{\rho_{m}}\Big)e^{\sqrt{3\lambda(1+\alpha)}\phi})\Big].\nonumber
\end{eqnarray}
 Again here $\lambda$ accounts for $\alpha<-1$ values.
 
We can distinguish three different cases:

{\em i)} {\bf $ {\rho}_\Lambda>0$ and $ {\rho}_m>0$}

\begin{figure}[ht!]
  \centering
    \includegraphics[width=8truecm]{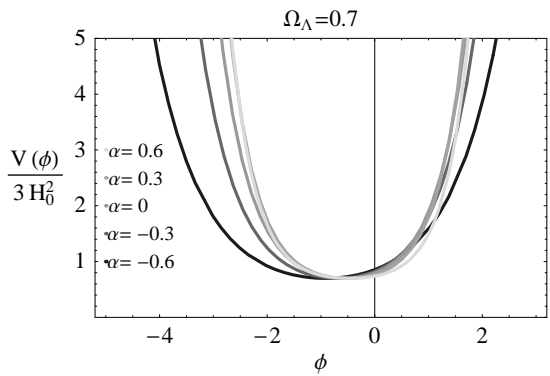}
    \caption{ \small \small Scalar field potential with $ {\rho}_\Lambda,\rho_{m}>0$, non phantom case.}
    \label{fig:pot1}
\end{figure}
The minimum of the potential here is clearly $V(\phi_{min})= {\rho}_\Lambda$ and the field rolls down to it, reaching the final de Sitter attractor (see Fig.~\ref{fig:pot1}).

{\em ii)} {\bf{$ {\rho}_\Lambda>0$ and $ {\rho}_m<0$, phantom}}

\begin{figure}[ht!]
  \centering
    \includegraphics[width=8truecm]{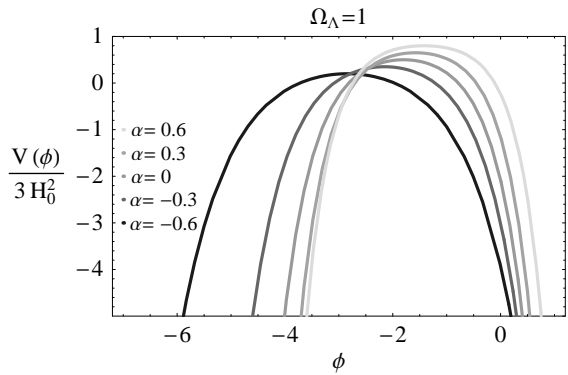}
    \caption{ \small \small Scalar field potential with $ {\rho}_\Lambda>0$ and  $\rho_{m}<0$, phantom case.}
    \label{fig:pot2}
\end{figure}
The universe evolves from a contracting phase, bounce at $a_{*}=(|\rho_{m}|/\rho_{\Lambda})^{\frac{1}{3(1+\alpha)}}$ and then re-expands \cite{Ananda:astro-ph/0512224}.
The potential here exhibits a maximum and the field is then forced to climb up the hill towards the maximum (see Fig.~\ref{fig:pot2}) reaching the ensuing De Sitter attractor. The scalar field thus exhibits phantom-like behaviour  with the energy density growing with time. 

{\em iii)} {\bf$ {\rho}_\Lambda<0$ and $ {\rho}_m>0$, non phantom}

\begin{figure}[ht!]
  \centering
    \includegraphics[width=8truecm]{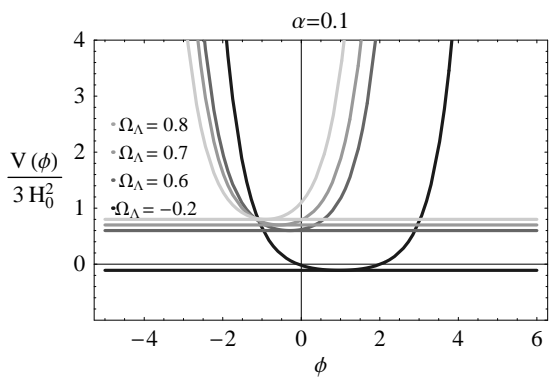}
    \caption{ \small \small Scalar field potential for different values of $ {\rho}_\Lambda$, non phantom case.  The minimum of the potential is ${\rho}_\Lambda (1+\alpha)/2$ for  $ {\rho}_\Lambda<0$.}
    \label{fig:pot3}
\end{figure}

The case $ {\rho}_\Lambda<0$ necessarly gives $ {\rho}_m>0$, if we assume a positive total energy density of the fluid. The universe expands to a maximum $a_{max}=(\rho_{m}/|\rho_{\Lambda}|)^{\frac{1}{3(1+\alpha)}}$ and then recollapses. At $a=a_{max}$ the field has reached the minimum of the potential, which in this case is not $ {\rho}_\Lambda$, but is $V(\phi_{min})= {\rho}_\Lambda (1+\alpha)/2$ (see  Fig.~\ref{fig:pot3} for negative value of $\Omega_{\Lambda}$). 

The potential (\ref{eq:vdip}) has been derived imposing an affine equation of state for a barotropic fluid. 
However, as said above, the scalar field 
 has an extra degree of freedom, hence the universe expansion in general will not be the same that in the barotropic fluid case. In order to study the dynamics of the scalar field we need to look at the Klein Gordon equation and its phase space. In Ref.~\cite{Ananda:astro-ph/0512224} it has been shown that current background observables exclude phantom behaviour at more than $3\sigma$. Moreover, as we will see, the de Sitter critical point in the phase space is a pure attractor only in non phantom cases. For these reasons till the end of this section we will put $\lambda=1$, $\alpha>-1$ and $\rho_{m}>0$.
 
Rescaling the scalar field so as to have the minimum at $\varphi=\phi-\phi_{min}=0$, the potential  (\ref{eq:vdip}) can be written as
\begin{equation}\label{eq:vdip2}
V(\varphi)= \rho_{\Lambda}\Big[\frac{3+\alpha}{4}+\frac{(1-\alpha)}{4}\cosh{(\varphi\sqrt{3(1+\alpha)})}\Big].
\end{equation}
A related form was derived in \cite{Gorini:hep-th/0311111}.
Here the case $\rho_{\Lambda}=0$  is not included, since we put $e^{-\sqrt{3(1+\alpha)}\phi_{min}}=\rho_{\Lambda}/\rho_{m}$, but it is recovered in the limit $\phi_{min}\rightarrow \infty$ (corresponding to the pure exponential potential, where $\alpha$ takes the standard role of the linear EoS parameter: $P_{x}=\alpha \rho_{x}$). Thus, assuming as a starting point the  affine EoS we fall into the class of quintessence ``exponential potentials'', whose properties are well known (\cite{Copeland:gr-qc/9711068,Ferreira:astro-ph/9711102,Liddle:astro-ph/9809272,Sahni:astro-ph/9910097}) and which we will comment more in details further on.

\begin{figure}[ht!]
  \centering
    \includegraphics[width=8truecm]{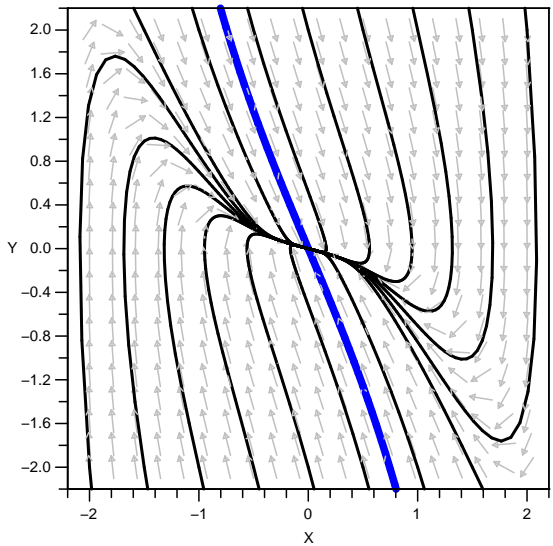}
    \caption{ \small \small Phase space for system (\ref{xdot})-(\ref{ydot}) with $\alpha=0.8$ and $\Omega_{\Lambda}=0.7$. The fixed point at the origin is a stable node for $\alpha\not =0$ and represents a de Sitter asymptotic state. The thicker line represents the curve (\ref{eq:track}), i.e. the solutions corresponding exactly to the affine EoS fluid. It is rather clear that generic trajectories approach the fixed point along the eigenvector corresponding to $e_2$ and not along the thick curve, see text.
}     \label{fig:track1}
\end{figure}

In order to analyse the behaviour  of the field from a dynamical system point of view,  let us define the new dimensionless variables
\begin{eqnarray}\label{eq:newv1}
X:=\varphi,\qquad
Y:=\frac{d\varphi}{d \eta},\qquad
\eta:=\sqrt{ {\rho}_m}t.
\end{eqnarray}
Then the Klein-Gordon and Friedmann equations are equivalent to the system
\begin{eqnarray}\label{xdot}
X'&=&Y,\\
Y'&=&-\sqrt{3} \Big(\frac{Y^2}{2}+\frac{V}{ {\rho}_m}\Big)^{1/2}Y-\frac{1}{ {\rho}_m}\frac{dV}{dX}, \label{ydot}
\end{eqnarray}
where the prime represents the derivative with respect to $\eta$.  System (\ref{xdot})-(\ref{ydot}) obviously exhibits a fixed point ($0,0$),  corresponding to the field lying at rest   at the minimum of the potential and driving the expansion with an effective EoS parameter (\ref{cs2})  $w_{\phi}=-1$, i.e.\ an effective cosmological constant. This point therefore represents a de Sitter model. The eigenvalues of the linearization of  system (\ref{xdot})-(\ref{ydot}) at this critical point are $e_1=-\sqrt{3\rho_{\Lambda}/{(4\rho_{m})}}(1-\alpha)$ and  $e_2=-\sqrt{3\rho_{\Lambda}/{(4\rho_{m})}}(1+\alpha)$, both always negative for $|\alpha|<1$ and $\rho_{\Lambda},\rho_{m}>0$.  Within these bounds, which we always assume,  this fixed point is therefore an attractor: a stable node in general, and an improper node in the degenerate case $e_1=e_2=-\sqrt{{3\rho_{\Lambda}}/{4}}$, i.e.\ for $\alpha=0$. Therefore, the scalar field forcely approaches asymptotically the de Sitter attractor $(0,0)$ in the phase space, which corresponds to the total domination of the constant part of the potential and the late time accelerated expansion (see Fig.~\ref{fig:track1}).

It is easy to derive  an analytic expression  in terms of $X$  and $Y$ for the trajectories corresponding to the affine EoS (\ref{EOSaffine}):  they lie on the curve 
\begin{equation}\label{eq:track}
Y_{aff}=\pm \sqrt{\frac{\rho_{\Lambda}}{\rho_{m}}\frac{(1+\alpha)}{2}(\cosh{(X_{aff}\sqrt{3(1+\alpha)})}-1)},
\end{equation}
which we may call the affine curve in phase space. This is the union of three exact solutions of system (\ref{xdot})-(\ref{ydot}): 1) the fixed point $(0,0)$; 2) a positive branch for $X<0$; 3) a negative branch for $X>0$.  
\begin{figure}[ht!]
  \centering
    \includegraphics[width=8truecm]{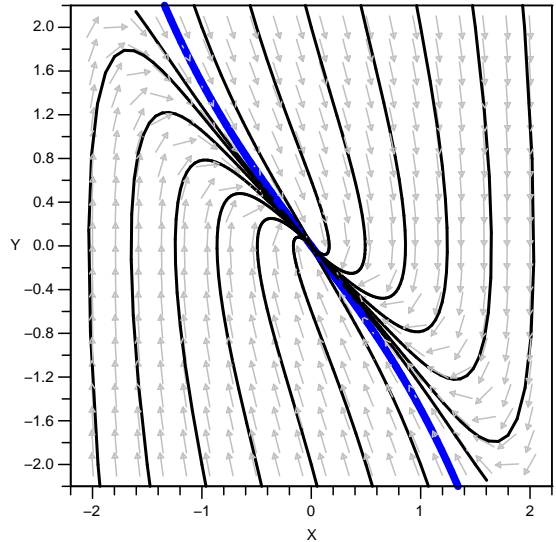}
    \caption{ \small \small Phase space for system (\ref{xdot})-(\ref{ydot}) with $\alpha=0$ and $\Omega_{\Lambda}=0.7$.  Here generic trajectories approach the fixed point along the single existing eigenvector $e_{1}=e_2$, see text.
}     \label{fig:track2}
\end{figure}
\begin{figure}[ht!]
  \centering
    \includegraphics[width=8truecm]{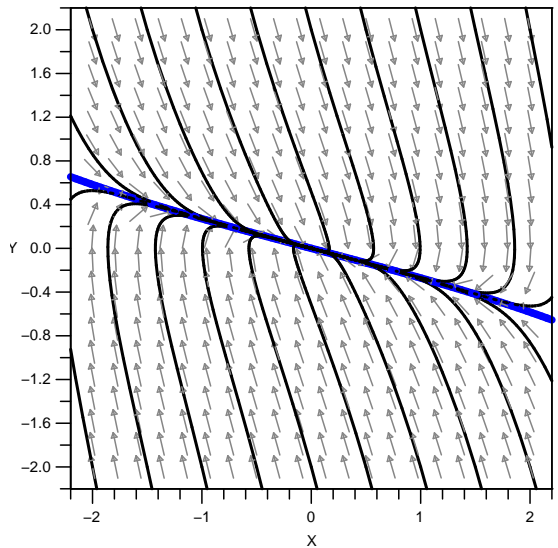}
    \caption{ \small \small Phase space for system (\ref{xdot})-(\ref{ydot}) with $\alpha=-0.8$ and $\Omega_{\Lambda}=0.7$. For negative values of $\alpha$ the qualitative behavior is rather clear: generic trajectories approach the fixed point along the eigenvector corresponding to $e_2$, driven towards the affine curve.
}     \label{fig:track3}
\end{figure}

An analysis of the linearization of system  (\ref{xdot})-(\ref{ydot}) shows  that, the eigenvector $\mathbf{E}_2$ corresponding to $e_2$ is tangent at the origin to the affine curve (\ref{eq:track}). For $\alpha>0$ $e_{2}<e_{1}<0$,  thus generic trajectories approach the  de Sitter stable node along the other eigenvector  $\mathbf{E}_1$ (they are tangent to it), as it is rather clear from the representative example in Fig. \ref{fig:track1}. For the improper node case $\alpha=0$ there is a single eigenvector, and all trajectories approach the stable node from that direction. In other words, for $\alpha > 0$ the scalar field dynamics mimics the affine EoS fluid  rather poorly, unless $\alpha$ is very small or zero.  In practice, for the potentially interesting case  $\alpha\ll 1$ \cite{Balbi:astro-ph/0702423,Muller:astro-ph/0410621,NEXT2}, the angle between the two eigenvectors becomes smaller and smaller, and the scalar field evolution is better and better represented by that of the affine fluid. The situation changes for for $\alpha < 0$: in this case $e_1<e_2<0$ and generic trajectories  approach the  de Sitter stable node along the  eigenvector  $\mathbf{E}_2$ (tangent to it), which remains tangent to the  affine curve (\ref{eq:track}). In this case therefore, the affine curve (\ref{eq:track}) attracts generic trajectories, better and better for $1+\alpha \rightarrow 0^+$ (see Fig. 5).

As analysed in Ref.~\cite{Sahni:astro-ph/9910097} the scalar field crosses two distinct phases: first of all it acts as in a standard pure exponential quintessence model, in which phase, if coexisting together with other components (e.g. radiation and CDM), it can track the EoS of the background  and, secondly,  it slips into the minimum following a parabolic-like potential. However, differently from Ref.~\cite{Sahni:astro-ph/9910097}, here the minimum is non-zero. Thus, when entering the parabolic phase an effective cosmological constant drives the accelerated expansion.

It is often asserted that, since the speed of  sound  for a scalar field is equal to the speed of light, then the scalar field cannot form sub-horizon structures. One should exercise extreme caution to relate the definition of the speed of sound to the evolution of density perturbations like in the hydrodynamic models.  Indeed it is true that in the perturbation equation of a scalar field the term which is proportional to the wave number $k^{2}$ and define the phase velocity is exactly the unity, and that this term is the well known sound of speed contribution to a standard wave equation, see Eq.~(\ref{cs2}). On the other hand, the second derivative of the potential with respect to the field plays an important role in the clustering. Let us write down the relativistic equation for the scalar perturbations:
\begin{equation}
\label{scalpert}
\ddot{\delta\varphi_{k}}+3H\dot{\delta\varphi_{k}}+\Big(\frac{k^{2}}{a^{2}}+V''(\varphi)\Big)\delta\varphi_{k}=-\frac{1}{2}\dot{\varphi}\dot{h},
\end{equation}
where the prime here indicates the derivative with respect to the field, $h$ is the trace of the spatial  metric tensor perturbation  in the synchronous gauge and $k=2\pi/\lambda$ is the wave number. 

At early time the potential (\ref{eq:vdip2}) is undistinguishable from a simple exponential potential and the energy density of the scalar field tracks the background radiation evolution \cite{Ferreira:astro-ph/9711102}. However, soon after the equivalence of matter and radiation, one would like  the scalar field to mimic a matter component. If the scalar field then leaves the exponential regime and enters the parabolic phase, its fast oscillations around the minimum could generate an average energy density which acts as a matter-like energy density, see e.g.\ \cite{Matos:astro-ph/0006024}.  The second derivative of the potential today is the square of the mass associated to the fluctuations ($m_{\varphi}^{2}$) and scalar perturbations can only grow if the term proportional to $k^{2}$ in (\ref{scalpert}) is subdominant with respect to $a^{2}V''$, i.e. for $k<am_{\varphi}$. This mass defines the value of the Compton length: 
\begin{equation}
\label{mass}
\lambda_{c}=\frac{2 \pi}{\sqrt{V''}}=\frac{4\pi}{\sqrt{3\rho_{\Lambda}(1-\alpha^{2})}},
\end{equation}
In the most optimistic view we can set the stage of oscillations to begin just before the equivalence, i.e. at $z\simeq 10^{5}$. In this case, if one thinks about a scalar field related to ultra-light particles capable to condensate and set an energy scale of the potential $V_{0}\simeq 10^{6}Mpc^{-2}$ \cite{Sahni:astro-ph/9910097,Matos:astro-ph/0006024}, a Compton length which is of order of hundreds parsec is obtained.  Besides, under the assumption of a unified dark matter suitable for driving also the late-time acceleration, this description is no longer valid. As already mentioned, after the early exponential phase, the potential enters a parabolic phase, where its shape is well represented by the quadratic expression $V(\varphi)\simeq \rho_{\Lambda}/4((3+\alpha)+(1-\alpha)(1+3(1+\alpha)\varphi^{2}))$. The mass associated to the fluctuations stabilizes when the frequency of the oscillations is bigger than the Hubble expansion, namely when $m_{\varphi}\ge H$, otherwise it varies until it asymptotically mimics the parabolic oscillations set by this potential, where $V''$, calculated in the minimum, is the value entering Eq.~(\ref{mass}). Indeed the energy scale related to the potential (\ref{eq:vdip2}) is $V_{0}=\rho_{\Lambda}\simeq 10^{-9}Mpc^{-2}$, for which we find a Compton length much larger than the horizon, even for tiny values of $\alpha$. Unavoidably, scalar perturbations at scales smaller than the horizon must have been erased, with unpleasant consequences for structure formation. Hence one is forced either to trigger two different energy scales in the potential, no longer describing an affine EoS, or to use the scalar field only as a dark energy component. Infact, when a scalar field is invoked merely as a dark energy component, a naturally vanishing mass is recovered. Moreover the homogeneity of the dark energy is a desirable property, as we do not observe its clustering at sub-horizon scales. 

 In the next subsection we will present a purely kinetic  Lagrangian which can avoid this problem.

\subsection{K-essence}
\label{sec_k}
The possibility of k-essence was first taken into consideration to explain the dynamics of inflation \cite{Armendariz-Picon:hep-th/9904075}, but has been subsequently proposed to generate the late time acceleration \cite{Chiba:astro-ph/9912463}. As already mentioned in the introduction, k-essence is particularly suitable for describing unified fluids with a matter-like component and a subluminal sound of speed. We now discuss a purely kinetic k-essence model equivalent to an affine EoS. Let us assume that the  Lagrangian is a generic function of $\chi$ only, i.e.
\begin{equation}\label{lag_k}
\mathcal{L}=P(\chi),
\end{equation}
which hides a simple constant term playing the role of a potential \cite{Giannakis:astro-ph/0501423v2}. More general types of  Lagrangian including, as a factor in (\ref{lag_k}), a function of the field can also be motivated by low energy effective string theory (see Ref.~\cite{Copeland:hep-th/0603057v3}). A general expression for the Lagrangian of a purely kinetic k-essence model obtained from a barotropic EoS was derived in \cite{feinstein:arXiv:gr-qc/0501101}.

In a Friedman-Robertson-Walker metric the equation of motion for the field is
\begin{equation}\label{eqmot}
\Big(\frac{d P}{d\chi}+2\chi\frac{d^{2}P}{d \chi^{2}}\Big)\chi'+6\chi \frac{d P}{d\chi}=0,
\end{equation}
where $'=d/d \log{(a)}$.
From this equation a straight indipendence of the solution on other possible components is easily noticeable. Moreover the EoS and the speed of sound  do not seem to be influenced by the choice of the constant potential, but they actually are, in that the potential affects the evolution of $\chi$ and, in turn, through the Friedman equation the other components.

Assuming a specific function $P(\chi)$ one can in principle solve Eq.~(\ref{eqmot}) for $\chi (a)$ and then substitute it in the expression (\ref{density}) for $\rho_{\phi}(\chi)$  to figure out the evolution of the energy density with the scale factor. In \cite{bertacca:astro-ph/0703259} the stable nodes of Eq.~(\ref{eqmot}) have been analysed, corresponding to solutions for which either $\partial P/\partial \chi|_{\chi_{*}}=0$ or $\chi_{*}=0$ (both with $w_{\phi}=-1$). Thus, modelling $P$ by means of a parabolic form, or equivalently expanding the pressure around its minimum in $\chi_{*}$ (as previously done in\cite{Scherrer:astro-ph/0402316}), they found $\rho_{\phi}=-g_{0}+4g_{2}\chi_{*}^{2}(a/a_{1})^{-3}$, where $g_{0}$ and $g_{2}$ are constants. This corresponds to an energy density of a unified dark fluid composed by a cosmological constant and a {\it pure} CDM. However this solution happens to be an approximation  only valid in the neighborhood of the minimum, and for $a\gg a_{1} $, where $a_{1}$ depends on the constants. Moreover the speed of sound is not vanishing at all out of this regime. Looking for an explicit solution with a $\chi^{2}$  Lagrangian leads to undesirable radiation-like evolution in the past. 
Here our approach moves along an inverse path: we start imposing the evolution for the energy density corresponding to the affine EoS (\ref{EOSaffine}) and derive the expression for $P(\chi)$ and the speed of sound.

Connecting Eq.(\ref{EOSaffine}) to Eq.~(\ref{density}), a first order differential equation in $P$ as a function of $\chi$ is derived:
\begin{equation}\label{pidichi}
P-2\alpha \chi \frac{dP}{d\chi}+\rho_{\Lambda}=0.
\end{equation}
This equation can be solved for the pressure and then used in Eq.~(\ref{density}) to derive the following:
\begin{equation}
\label{pidichi2}
P=-\rho_{\Lambda}+c\chi^{\frac{1+\alpha}{2\alpha}};\qquad \rho_{\phi}=\rho_{\Lambda}+\frac{c}{\alpha}\chi^{\frac{1+\alpha}{2\alpha}},
\end{equation}
where $c=\rho_{m}\alpha/\chi_{0}^{\frac{1+\alpha}{2\alpha}}$ is the integration constant derived imposing the value of the fluid energy density at present and $\chi_{0}$ is $\chi$ at present time.  These two function are clearly solutions of the affine EoS (\ref{EOSaffine}), and the stable node of Eq.~(\ref{eqmot}) is the one for which $\chi_{0}=0$ and $w_{\phi}=-1$.  The expressions for the pressure and the energy density (\ref{pidichi2}) represent the sum of a power-law solution, related to an unchanging and non-zero EoS ($p=\alpha\rho$), plus a constant potential term that acts as a cosmological constant \cite{Tejedor:gr-qc/0604031}. It has been shown in  \cite{Chimento:astro-ph/0305007} that if $\alpha\ge 0$ the constant EoS is an attractor for the k-essence field, holding also for superluminal EoS. From Eq.~(\ref{cs2}) it follows that the ``effective'' speed of sound  in this model is $c_{s}^{2}=\alpha$. In many k-essence model, whenever $\partial^{2} P/\partial \chi^{2}\neq 0$ and positive, the speed of sound can be arbitrarily small between last scattering and today. As a purely kinetic k-essence model, one main property is that the scalar field possesses a single degree of freedom: therefore its EoS is naturally barotropic and reads:
\begin{equation}
\label{wk}
w_{\phi}=\frac{-\rho_{\Lambda}a^{3(1+\alpha)}+\rho_{m}\alpha}{\rho_{\Lambda}a^{3(1+\alpha)}+\rho_{m}}.
\end{equation}
Hence the scalar field dynamics 
derived from a  purely kinetic  Lagrangian, related to the simplest first order parameterization of a perfect fluid EoS, is completely equivalent to the perfect fluid description: while $-1\leq w_{\phi}\leq\alpha$, the speed of sound is constant and can reproduce a ''fuzzy'' dark matter \cite{Hu:astro-ph/0003365}, i.e.  a low sound-speed fluid inhibiting  growth of matter inhomogeneities at very small scales. The density and pressure perturbations of the constant part of the  Lagrangian are null independently of the speed of sound, and the pressure perturbations are defined in the rest frame of the matter-like component. Moreover, in this case, the scalar field does not have the problem of an early radiation-like behavior. In addition, evolution (\ref{rhodia}) is not a transient solution, but is valid at all times, implying no unnatural fine-tuning on the parameters (apart from the energy scale). Besides, the effective speed of sound defines the stabilization scale for a perturbation \cite{Garriga:hep-th/9904176}; in this model perturbations of physical wavelength $a/k<c_{s}H_{0}^{-1}$ must have been completely erased. Thus, the Compton length related to the growth of inhomogeneities is $\lambda_{c}=\alpha H_{0}^{-1}$ and, in order to obtain formed structures at least at Kpc scale, we expect $\alpha \lesssim 10^{-6}$, i.e. very strongly constrained, although non-vanishing.  This is also the bound obtained in \cite{Muller:astro-ph/0410621} for a generalized dark matter from a direct comparison with the matter power spectrum. A larger value of $\alpha$ would indeed distort also the large scale CMB anisotropy power spectrum via the integrated Sachs Wolfe effect \cite{NEXT2}.
\\

\section{Conclusions}

We have investigated the prospects of reproducing the behaviour of a dark fluid with affine barotropic EoS within the context of a scalar field model with an appropriate Lagrangian. The importance of this approach is twofold: first, this kind of EoS can be regarded as a first order approximation to that of any generic barotropic fluid; second, describing this kind of fluid in a scalar field framework gives a more fundamental representation of the dark component. Our main finding is that it is indeed possible to obtain an equivalent description of the dark fluid with affine EoS in two cases: a quintessence-like Lagrangian, with a hyperbolic potential, and a purely kinetic Lagrangian, or k-essence. Both models are able to reproduce the correct  behaviour of the cosmological background from early epochs up to the present, although in the quintessence case this is strictly true only for $\alpha\leq 0$.

In the case of the quintessence potential we have found (see also \cite{Gorini:hep-th/0311111}), the dynamics of the scalar is driven by an ''exponential-like'' potential including a constant term. The field then first follow a trajectory where it tracks the dominant cosmological component (radiation or matter), then it reaches an attractor solution where it behaves as a cosmological constant, resulting in an accelerated expansion of the universe. However, when one addressed the problem from the point of view of structure formation, this kind of Lagrangian does not appear to have the necessary characteristics to act as a unified dark component, although it could be used as a dark energy. In fact, since for quintessence scalar field the speed of sound is such that $c_s^2=1$ and the energy scale related to the minimum is set by the cosmological constant, perturbations are erased on scales smaller than a Compton length potentially bigger than the horizon. The only way to use the scalar field as a unified dark fluid would  be to introduce a new energy scale in the potential. In this regard, the dark fluid obtained from a quintessence potential has a worse behaviour than the barotropic fluid that it tries to reproduce, since in the latter case the speed of sound can be made arbitrary small.

The k-essence Lagrangian seems more suitable to treat a unified dark fluid, even when considerations based on structure formation are taken into account. We found that the energy density scaling behavior of an affine perfect fluid is an exact solution of a specific pure kinetic Lagrangian. Furthermore, the absence of an early phase mimicking the radiation component makes the model suitable to reproduce a matter-like behaviour at all times. Moreover, in this case the speed of sound is related to one of the parameters of the affine EoS, $c_s^2=\alpha$. This suggests that the model might have the right clustering properties both at galactic and cosmological scale (although requiring a fine-tuning of the $\alpha$ parameter). Based on simple arguments related to the expected Compton length, we conclude that the $\alpha$ parameters should be strongly constrained to small ($\sim 10^{-6}$) but not necessarily vanishing values. This is consistent with values obtained by a comparison of a similar model (namely, a generalized dark matter plus a cosmological constant) with the matter power spectrum\cite{Muller:astro-ph/0410621}. A full analysis in linear perturbation theory and a comparison of the predictions with CMB anisotropy data is currently being undertaken and will be the subject of a forthcoming paper \cite{NEXT2}.

\subsection{Acknowledgements} 
We would like to thank Luca Amendola and Davide Pietrobon for useful discussions about the topics.

\bibliography{Bscal}

\begin{thebibliography}{41}
\expandafter\ifx\csname natexlab\endcsname\relax\def\natexlab#1{#1}\fi
\expandafter\ifx\csname bibnamefont\endcsname\relax
  \def\bibnamefont#1{#1}\fi
\expandafter\ifx\csname bibfnamefont\endcsname\relax
  \def\bibfnamefont#1{#1}\fi
\expandafter\ifx\csname citenamefont\endcsname\relax
  \def\citenamefont#1{#1}\fi
\expandafter\ifx\csname url\endcsname\relax
  \def\url#1{\texttt{#1}}\fi
\expandafter\ifx\csname urlprefix\endcsname\relax\def\urlprefix{URL }\fi
\providecommand{\bibinfo}[2]{#2}
\providecommand{\eprint}[2][]{\url{#2}}

\bibitem[{\citenamefont{{Spergel} et~al.}(2006)\citenamefont{{Spergel}, {Bean},
  {Dor{\'e}}, {Nolta}, {Bennett}, {Dunkley}, {Hinshaw}, {Jarosik}, {Komatsu},
  {Page} et~al.}}]{Spergel:astro-ph/0603449}
\bibinfo{author}{\bibfnamefont{D.~N.} \bibnamefont{{Spergel}}},
  \bibinfo{author}{\bibfnamefont{R.}~\bibnamefont{{Bean}}},
  \bibinfo{author}{\bibfnamefont{O.}~\bibnamefont{{Dor{\'e}}}},
  \bibinfo{author}{\bibfnamefont{M.~R.} \bibnamefont{{Nolta}}},
  \bibinfo{author}{\bibfnamefont{C.~L.} \bibnamefont{{Bennett}}},
  \bibinfo{author}{\bibfnamefont{J.}~\bibnamefont{{Dunkley}}},
  \bibinfo{author}{\bibfnamefont{G.}~\bibnamefont{{Hinshaw}}},
  \bibinfo{author}{\bibfnamefont{N.}~\bibnamefont{{Jarosik}}},
  \bibinfo{author}{\bibfnamefont{E.}~\bibnamefont{{Komatsu}}},
  \bibinfo{author}{\bibfnamefont{L.}~\bibnamefont{{Page}}},
  \bibnamefont{et~al.}, \bibinfo{journal}{ArXiv Astrophysics e-prints}
  (\bibinfo{year}{2006}), \eprint{astro-ph/0603449}.

\bibitem[{\citenamefont{{Springel} et~al.}(2006)\citenamefont{{Springel},
  {Frenk}, and {White}}}]{2006Natur.440.1137S}
\bibinfo{author}{\bibfnamefont{V.}~\bibnamefont{{Springel}}},
  \bibinfo{author}{\bibfnamefont{C.~S.} \bibnamefont{{Frenk}}},
  \bibnamefont{and} \bibinfo{author}{\bibfnamefont{S.~D.~M.}
  \bibnamefont{{White}}}, \bibinfo{journal}{\nat}
  \textbf{\bibinfo{volume}{440}}, \bibinfo{pages}{1137} (\bibinfo{year}{2006}),
  \eprint{arXiv:astro-ph/0604561}.

\bibitem[{\citenamefont{{Perlmutter} et~al.}(1999)\citenamefont{{Perlmutter},
  {Aldering}, {Goldhaber}, {Knop}, {Nugent}, {Castro}, {Deustua}, {Fabbro},
  {Goobar}, {Groom} et~al.}}]{Perlmutter:astro-ph/9812133}
\bibinfo{author}{\bibfnamefont{S.}~\bibnamefont{{Perlmutter}}},
  \bibinfo{author}{\bibfnamefont{G.}~\bibnamefont{{Aldering}}},
  \bibinfo{author}{\bibfnamefont{G.}~\bibnamefont{{Goldhaber}}},
  \bibinfo{author}{\bibfnamefont{R.~A.} \bibnamefont{{Knop}}},
  \bibinfo{author}{\bibfnamefont{P.}~\bibnamefont{{Nugent}}},
  \bibinfo{author}{\bibfnamefont{P.~G.} \bibnamefont{{Castro}}},
  \bibinfo{author}{\bibfnamefont{S.}~\bibnamefont{{Deustua}}},
  \bibinfo{author}{\bibfnamefont{S.}~\bibnamefont{{Fabbro}}},
  \bibinfo{author}{\bibfnamefont{A.}~\bibnamefont{{Goobar}}},
  \bibinfo{author}{\bibfnamefont{D.~E.} \bibnamefont{{Groom}}},
  \bibnamefont{et~al.}, \bibinfo{journal}{\apj} \textbf{\bibinfo{volume}{517}},
  \bibinfo{pages}{565} (\bibinfo{year}{1999}), \eprint{arXiv:astro-ph/9812133}.

\bibitem[{\citenamefont{{Riess} et~al.}(2007)\citenamefont{{Riess}, {Strolger},
  {Casertano}, {Ferguson}, {Mobasher}, {Gold}, {Challis}, {Filippenko}, {Jha},
  {Li} et~al.}}]{Riess:astro-ph/0611572}
\bibinfo{author}{\bibfnamefont{A.~G.} \bibnamefont{{Riess}}},
  \bibinfo{author}{\bibfnamefont{L.-G.} \bibnamefont{{Strolger}}},
  \bibinfo{author}{\bibfnamefont{S.}~\bibnamefont{{Casertano}}},
  \bibinfo{author}{\bibfnamefont{H.~C.} \bibnamefont{{Ferguson}}},
  \bibinfo{author}{\bibfnamefont{B.}~\bibnamefont{{Mobasher}}},
  \bibinfo{author}{\bibfnamefont{B.}~\bibnamefont{{Gold}}},
  \bibinfo{author}{\bibfnamefont{P.~J.} \bibnamefont{{Challis}}},
  \bibinfo{author}{\bibfnamefont{A.~V.} \bibnamefont{{Filippenko}}},
  \bibinfo{author}{\bibfnamefont{S.}~\bibnamefont{{Jha}}},
  \bibinfo{author}{\bibfnamefont{W.}~\bibnamefont{{Li}}}, \bibnamefont{et~al.},
  \bibinfo{journal}{\apj} \textbf{\bibinfo{volume}{659}}, \bibinfo{pages}{98}
  (\bibinfo{year}{2007}), \eprint{arXiv:astro-ph/0611572}.

\bibitem[{\citenamefont{Visser}(1997)}]{Visser:1997aa}
\bibinfo{author}{\bibfnamefont{M.}~\bibnamefont{Visser}},
  \bibinfo{journal}{Science} \textbf{\bibinfo{volume}{276}},
  \bibinfo{pages}{88} (\bibinfo{year}{1997}).

\bibitem[{\citenamefont{{Copeland} et~al.}(2006)\citenamefont{{Copeland},
  {Sami}, and {Tsujikawa}}}]{Copeland:hep-th/0603057v3}
\bibinfo{author}{\bibfnamefont{E.~J.} \bibnamefont{{Copeland}}},
  \bibinfo{author}{\bibfnamefont{M.}~\bibnamefont{{Sami}}}, \bibnamefont{and}
  \bibinfo{author}{\bibfnamefont{S.}~\bibnamefont{{Tsujikawa}}},
  \bibinfo{journal}{ArXiv High Energy Physics - Theory e-prints}
  (\bibinfo{year}{2006}).

\bibitem[{\citenamefont{{Kamenshchik} et~al.}(2001)\citenamefont{{Kamenshchik},
  {Moschella}, and {Pasquier}}}]{Kamenshchik:gr-qc/0103004}
\bibinfo{author}{\bibfnamefont{A.}~\bibnamefont{{Kamenshchik}}},
  \bibinfo{author}{\bibfnamefont{U.}~\bibnamefont{{Moschella}}},
  \bibnamefont{and}
  \bibinfo{author}{\bibfnamefont{V.}~\bibnamefont{{Pasquier}}},
  \bibinfo{journal}{Physics Letters B} \textbf{\bibinfo{volume}{511}},
  \bibinfo{pages}{265} (\bibinfo{year}{2001}), \eprint{arXiv:gr-qc/0103004}.

\bibitem[{\citenamefont{{Bento} et~al.}(2004)\citenamefont{{Bento},
  {Bertolami}, and {Sen}}}]{Bento:astro-ph/0407239}
\bibinfo{author}{\bibfnamefont{M.~C.} \bibnamefont{{Bento}}},
  \bibinfo{author}{\bibfnamefont{O.}~\bibnamefont{{Bertolami}}},
  \bibnamefont{and} \bibinfo{author}{\bibfnamefont{A.~A.} \bibnamefont{{Sen}}},
  \bibinfo{journal}{\prd} \textbf{\bibinfo{volume}{70}},
  \bibinfo{pages}{083519} (\bibinfo{year}{2004}),
  \eprint{arXiv:astro-ph/0407239}.

\bibitem[{\citenamefont{{Balbi} et~al.}(2007)\citenamefont{{Balbi}, {Bruni},
  and {Quercellini}}}]{Balbi:astro-ph/0702423}
\bibinfo{author}{\bibfnamefont{A.}~\bibnamefont{{Balbi}}},
  \bibinfo{author}{\bibfnamefont{M.}~\bibnamefont{{Bruni}}}, \bibnamefont{and}
  \bibinfo{author}{\bibfnamefont{C.}~\bibnamefont{{Quercellini}}},
  \bibinfo{journal}{ArXiv Astrophysics e-prints}  (\bibinfo{year}{2007}),
  \eprint{astro-ph/0702423}.

\bibitem[{\citenamefont{{Amendola} et~al.}(2003)\citenamefont{{Amendola},
  {Finelli}, {Burigana}, and {Carturan}}}]{Amendola:astro-ph/0304325}
\bibinfo{author}{\bibfnamefont{L.}~\bibnamefont{{Amendola}}},
  \bibinfo{author}{\bibfnamefont{F.}~\bibnamefont{{Finelli}}},
  \bibinfo{author}{\bibfnamefont{C.}~\bibnamefont{{Burigana}}},
  \bibnamefont{and}
  \bibinfo{author}{\bibfnamefont{D.}~\bibnamefont{{Carturan}}},
  \bibinfo{journal}{Journal of Cosmology and Astro-Particle Physics}
  \textbf{\bibinfo{volume}{7}}, \bibinfo{pages}{5} (\bibinfo{year}{2003}),
  \eprint{arXiv:astro-ph/0304325}.

\bibitem[{\citenamefont{{Sahni} and {Wang}}(2000)}]{Sahni:astro-ph/9910097}
\bibinfo{author}{\bibfnamefont{V.}~\bibnamefont{{Sahni}}} \bibnamefont{and}
  \bibinfo{author}{\bibfnamefont{L.}~\bibnamefont{{Wang}}},
  \bibinfo{journal}{\prd} \textbf{\bibinfo{volume}{62}},
  \bibinfo{pages}{103517} (\bibinfo{year}{2000}),
  \eprint{arXiv:astro-ph/9910097}.

\bibitem[{\citenamefont{{Mainini} et~al.}(2005)\citenamefont{{Mainini},
  {Colombo}, and {Bonometto}}}]{Mainini:astro-ph/0503036v2}
\bibinfo{author}{\bibfnamefont{R.}~\bibnamefont{{Mainini}}},
  \bibinfo{author}{\bibfnamefont{L.~P.~L.} \bibnamefont{{Colombo}}},
  \bibnamefont{and} \bibinfo{author}{\bibfnamefont{S.~A.}
  \bibnamefont{{Bonometto}}}, \bibinfo{journal}{\apj}
  \textbf{\bibinfo{volume}{632}}, \bibinfo{pages}{691} (\bibinfo{year}{2005}).

\bibitem[{\citenamefont{{Giannakis} and
  {Hu}}(2005)}]{Giannakis:astro-ph/0501423v2}
\bibinfo{author}{\bibfnamefont{D.}~\bibnamefont{{Giannakis}}} \bibnamefont{and}
  \bibinfo{author}{\bibfnamefont{W.}~\bibnamefont{{Hu}}},
  \bibinfo{journal}{\prd} \textbf{\bibinfo{volume}{72}},
  \bibinfo{pages}{063502} (\bibinfo{year}{2005}).

\bibitem[{\citenamefont{{Be{\c c}a} and {Avelino}}(2007)}]{Beca:astro-ph/}
\bibinfo{author}{\bibfnamefont{L.~M.~G.} \bibnamefont{{Be{\c c}a}}}
  \bibnamefont{and} \bibinfo{author}{\bibfnamefont{P.~P.}
  \bibnamefont{{Avelino}}}, \bibinfo{journal}{MNRAS}
  \textbf{\bibinfo{volume}{376}}, \bibinfo{pages}{1169} (\bibinfo{year}{2007}).

\bibitem[{\citenamefont{{Chimento}}(2004)}]{Chimento:astro-ph/0311613}
\bibinfo{author}{\bibfnamefont{L.~P.} \bibnamefont{{Chimento}}},
  \bibinfo{journal}{\prd} \textbf{\bibinfo{volume}{69}},
  \bibinfo{pages}{123517} (\bibinfo{year}{2004}),
  \eprint{arXiv:astro-ph/0311613}.

\bibitem[{\citenamefont{{Scherrer}}(2004)}]{Scherrer:astro-ph/0402316}
\bibinfo{author}{\bibfnamefont{R.~J.} \bibnamefont{{Scherrer}}},
  \bibinfo{journal}{Physical Review Letters} \textbf{\bibinfo{volume}{93}},
  \bibinfo{pages}{011301} (\bibinfo{year}{2004}),
  \eprint{arXiv:astro-ph/0402316}.

\bibitem[{\citenamefont{{Bertacca} et~al.}(2007)\citenamefont{{Bertacca},
  {Matarrese}, and {Pietroni}}}]{bertacca:astro-ph/0703259}
\bibinfo{author}{\bibfnamefont{D.}~\bibnamefont{{Bertacca}}},
  \bibinfo{author}{\bibfnamefont{S.}~\bibnamefont{{Matarrese}}},
  \bibnamefont{and}
  \bibinfo{author}{\bibfnamefont{M.}~\bibnamefont{{Pietroni}}},
  \bibinfo{journal}{ArXiv Astrophysics e-prints}  (\bibinfo{year}{2007}),
  \eprint{astro-ph/0703259}.

\bibitem[{\citenamefont{{Visser}}(2004)}]{Visser:gr-qc/0309109}
\bibinfo{author}{\bibfnamefont{M.}~\bibnamefont{{Visser}}},
  \bibinfo{journal}{Classical and Quantum Gravity}
  \textbf{\bibinfo{volume}{21}}, \bibinfo{pages}{2603} (\bibinfo{year}{2004}),
  \eprint{arXiv:gr-qc/0309109}.

\bibitem[{\citenamefont{{Chiba} et~al.}(1997)\citenamefont{{Chiba}, {Sugiyama},
  and {Nakamura}}}]{Chiba:astro-ph/9704199}
\bibinfo{author}{\bibfnamefont{T.}~\bibnamefont{{Chiba}}},
  \bibinfo{author}{\bibfnamefont{N.}~\bibnamefont{{Sugiyama}}},
  \bibnamefont{and}
  \bibinfo{author}{\bibfnamefont{T.}~\bibnamefont{{Nakamura}}},
  \bibinfo{journal}{MNRAS} \textbf{\bibinfo{volume}{289}}, \bibinfo{pages}{L5}
  (\bibinfo{year}{1997}), \eprint{arXiv:astro-ph/9704199}.

\bibitem[{\citenamefont{Gorini et~al.}(2004)\citenamefont{Gorini, Kamenshchik,
  Moschella, and Pasquier}}]{Gorini:hep-th/0311111}
\bibinfo{author}{\bibfnamefont{V.}~\bibnamefont{Gorini}},
  \bibinfo{author}{\bibfnamefont{A.~Y.} \bibnamefont{Kamenshchik}},
  \bibinfo{author}{\bibfnamefont{U.}~\bibnamefont{Moschella}},
  \bibnamefont{and} \bibinfo{author}{\bibfnamefont{V.}~\bibnamefont{Pasquier}},
  \bibinfo{journal}{Phys. Rev.} \textbf{\bibinfo{volume}{D69}},
  \bibinfo{pages}{123512} (\bibinfo{year}{2004}), \eprint{hep-th/0311111}.

\bibitem[{\citenamefont{{Ananda} and {Bruni}}(2006)}]{Ananda:astro-ph/0512224}
\bibinfo{author}{\bibfnamefont{K.~N.} \bibnamefont{{Ananda}}} \bibnamefont{and}
  \bibinfo{author}{\bibfnamefont{M.}~\bibnamefont{{Bruni}}},
  \bibinfo{journal}{\prd} \textbf{\bibinfo{volume}{74}},
  \bibinfo{pages}{023523} (\bibinfo{year}{2006}),
  \eprint{arXiv:astro-ph/0512224}.

\bibitem[{\citenamefont{{Babichev} et~al.}(2005)\citenamefont{{Babichev},
  {Dokuchaev}, and {Eroshenko}}}]{babichev:arXiv:astro-ph/0407190}
\bibinfo{author}{\bibfnamefont{E.}~\bibnamefont{{Babichev}}},
  \bibinfo{author}{\bibfnamefont{V.}~\bibnamefont{{Dokuchaev}}},
  \bibnamefont{and}
  \bibinfo{author}{\bibfnamefont{Y.}~\bibnamefont{{Eroshenko}}},
  \bibinfo{journal}{Classical and Quantum Gravity}
  \textbf{\bibinfo{volume}{22}}, \bibinfo{pages}{143} (\bibinfo{year}{2005}),
  \eprint{arXiv:astro-ph/0407190}.

\bibitem[{\citenamefont{{Holman} and {Naidu}}(2004)}]{holman:astro-ph/0408102}
\bibinfo{author}{\bibfnamefont{R.}~\bibnamefont{{Holman}}} \bibnamefont{and}
  \bibinfo{author}{\bibfnamefont{S.}~\bibnamefont{{Naidu}}},
  \bibinfo{journal}{ArXiv Astrophysics e-prints}  (\bibinfo{year}{2004}),
  \eprint{astro-ph/0408102}.

\bibitem[{\citenamefont{{Hu}}(1998)}]{Hu:astro-ph/9801234}
\bibinfo{author}{\bibfnamefont{W.}~\bibnamefont{{Hu}}}, \bibinfo{journal}{\apj}
  \textbf{\bibinfo{volume}{506}}, \bibinfo{pages}{485} (\bibinfo{year}{1998}),
  \eprint{arXiv:astro-ph/9801234}.

\bibitem[{\citenamefont{Wald}(1983)}]{Wald:1983ky}
\bibinfo{author}{\bibfnamefont{R.~W.} \bibnamefont{Wald}},
  \bibinfo{journal}{Phys. Rev.} \textbf{\bibinfo{volume}{D28}},
  \bibinfo{pages}{2118} (\bibinfo{year}{1983}).

\bibitem[{\citenamefont{{Wainwright} and {Ellis}}(2005)}]{Wainwright:2005}
\bibinfo{author}{\bibfnamefont{J.}~\bibnamefont{{Wainwright}}}
  \bibnamefont{and} \bibinfo{author}{\bibfnamefont{G.~F.~R.}
  \bibnamefont{{Ellis}}}, \emph{\bibinfo{title}{{Dynamical Systems in
  Cosmology}}} (\bibinfo{publisher}{Dynamical Systems in Cosmology, Edited by
  J.~Wainwright and G.~F.~R.~Ellis, pp.~357.~ISBN 0521673526.~Cambridge, UK:
  Cambridge University Press, June 2005.}, \bibinfo{year}{2005}).

\bibitem[{\citenamefont{{Ellis} and {Madsen}}(1991)}]{1991CQGra...8..667E}
\bibinfo{author}{\bibfnamefont{G.~F.~R.} \bibnamefont{{Ellis}}}
  \bibnamefont{and} \bibinfo{author}{\bibfnamefont{M.~S.}
  \bibnamefont{{Madsen}}}, \bibinfo{journal}{Classical and Quantum Gravity}
  \textbf{\bibinfo{volume}{8}}, \bibinfo{pages}{667} (\bibinfo{year}{1991}).

\bibitem[{\citenamefont{{Garriga} and
  {Mukhanov}}(1999)}]{Garriga:hep-th/9904176}
\bibinfo{author}{\bibfnamefont{J.}~\bibnamefont{{Garriga}}} \bibnamefont{and}
  \bibinfo{author}{\bibfnamefont{V.~F.} \bibnamefont{{Mukhanov}}},
  \bibinfo{journal}{Physics Letters B} \textbf{\bibinfo{volume}{458}},
  \bibinfo{pages}{219} (\bibinfo{year}{1999}), \eprint{arXiv:hep-th/9904176}.

\bibitem[{\citenamefont{{Seljak} et~al.}(2006)\citenamefont{{Seljak}, {Slosar},
  and {McDonald}}}]{Seljak:astro-ph/0604335}
\bibinfo{author}{\bibfnamefont{U.}~\bibnamefont{{Seljak}}},
  \bibinfo{author}{\bibfnamefont{A.}~\bibnamefont{{Slosar}}}, \bibnamefont{and}
  \bibinfo{author}{\bibfnamefont{P.}~\bibnamefont{{McDonald}}},
  \bibinfo{journal}{Journal of Cosmology and Astro-Particle Physics}
  \textbf{\bibinfo{volume}{10}}, \bibinfo{pages}{14} (\bibinfo{year}{2006}),
  \eprint{arXiv:astro-ph/0604335}.

\bibitem[{\citenamefont{{Copeland} et~al.}(1998)\citenamefont{{Copeland},
  {Liddle}, and {Wands}}}]{Copeland:gr-qc/9711068}
\bibinfo{author}{\bibfnamefont{E.~J.} \bibnamefont{{Copeland}}},
  \bibinfo{author}{\bibfnamefont{A.~R.} \bibnamefont{{Liddle}}},
  \bibnamefont{and} \bibinfo{author}{\bibfnamefont{D.}~\bibnamefont{{Wands}}},
  \bibinfo{journal}{\prd} \textbf{\bibinfo{volume}{57}}, \bibinfo{pages}{4686}
  (\bibinfo{year}{1998}), \eprint{arXiv:gr-qc/9711068}.

\bibitem[{\citenamefont{{Ferreira} and
  {Joyce}}(1998)}]{Ferreira:astro-ph/9711102}
\bibinfo{author}{\bibfnamefont{P.~G.} \bibnamefont{{Ferreira}}}
  \bibnamefont{and} \bibinfo{author}{\bibfnamefont{M.}~\bibnamefont{{Joyce}}},
  \bibinfo{journal}{\prd} \textbf{\bibinfo{volume}{58}},
  \bibinfo{pages}{023503} (\bibinfo{year}{1998}),
  \eprint{arXiv:astro-ph/9711102}.

\bibitem[{\citenamefont{{Liddle} and
  {Scherrer}}(1999)}]{Liddle:astro-ph/9809272}
\bibinfo{author}{\bibfnamefont{A.~R.} \bibnamefont{{Liddle}}} \bibnamefont{and}
  \bibinfo{author}{\bibfnamefont{R.~J.} \bibnamefont{{Scherrer}}},
  \bibinfo{journal}{\prd} \textbf{\bibinfo{volume}{59}},
  \bibinfo{pages}{023509} (\bibinfo{year}{1999}),
  \eprint{arXiv:astro-ph/9809272}.

\bibitem[{\citenamefont{Muller}(2005)}]{Muller:astro-ph/0410621}
\bibinfo{author}{\bibfnamefont{C.~M.} \bibnamefont{Muller}},
  \bibinfo{journal}{Phys. Rev.} \textbf{\bibinfo{volume}{D71}},
  \bibinfo{pages}{047302} (\bibinfo{year}{2005}), \eprint{astro-ph/0410621}.

\bibitem[{\citenamefont{D.{Pietrobon et al.}}(2007)}]{NEXT2}
\bibinfo{author}{\bibnamefont{D.{Pietrobon et al.}}}, \bibinfo{journal}{{\it in
  preparation}}  (\bibinfo{year}{2007}).

\bibitem[{\citenamefont{{Matos} and {Arturo
  Ure{\~n}a-L{\'o}pez}}(2001)}]{Matos:astro-ph/0006024}
\bibinfo{author}{\bibfnamefont{T.}~\bibnamefont{{Matos}}} \bibnamefont{and}
  \bibinfo{author}{\bibfnamefont{L.}~\bibnamefont{{Arturo
  Ure{\~n}a-L{\'o}pez}}}, \bibinfo{journal}{\prd}
  \textbf{\bibinfo{volume}{63}}, \bibinfo{pages}{063506}
  (\bibinfo{year}{2001}), \eprint{arXiv:astro-ph/0006024}.

\bibitem[{\citenamefont{{Armend{\'a}riz-Pic{\'o}n}
  et~al.}(1999)\citenamefont{{Armend{\'a}riz-Pic{\'o}n}, {Damour}, and
  {Mukhanov}}}]{Armendariz-Picon:hep-th/9904075}
\bibinfo{author}{\bibfnamefont{C.}~\bibnamefont{{Armend{\'a}riz-Pic{\'o}n}}},
  \bibinfo{author}{\bibfnamefont{T.}~\bibnamefont{{Damour}}}, \bibnamefont{and}
  \bibinfo{author}{\bibfnamefont{V.}~\bibnamefont{{Mukhanov}}},
  \bibinfo{journal}{Physics Letters B} \textbf{\bibinfo{volume}{458}},
  \bibinfo{pages}{209} (\bibinfo{year}{1999}), \eprint{arXiv:hep-th/9904075}.

\bibitem[{\citenamefont{{Chiba} et~al.}(2000)\citenamefont{{Chiba}, {Okabe},
  and {Yamaguchi}}}]{Chiba:astro-ph/9912463}
\bibinfo{author}{\bibfnamefont{T.}~\bibnamefont{{Chiba}}},
  \bibinfo{author}{\bibfnamefont{T.}~\bibnamefont{{Okabe}}}, \bibnamefont{and}
  \bibinfo{author}{\bibfnamefont{M.}~\bibnamefont{{Yamaguchi}}},
  \bibinfo{journal}{\prd} \textbf{\bibinfo{volume}{62}},
  \bibinfo{pages}{023511} (\bibinfo{year}{2000}),
  \eprint{arXiv:astro-ph/9912463}.

\bibitem[{\citenamefont{{Diez-Tejedor} and
  {Feinstein}}(2005)}]{feinstein:arXiv:gr-qc/0501101}
\bibinfo{author}{\bibfnamefont{A.}~\bibnamefont{{Diez-Tejedor}}}
  \bibnamefont{and}
  \bibinfo{author}{\bibfnamefont{A.}~\bibnamefont{{Feinstein}}},
  \bibinfo{journal}{International Journal of Modern Physics D}
  \textbf{\bibinfo{volume}{14}}, \bibinfo{pages}{1561} (\bibinfo{year}{2005}),
  \eprint{arXiv:gr-qc/0501101}.

\bibitem[{\citenamefont{{D{\'{\i}}ez-Tejedor} and
  {Feinstein}}(2006)}]{Tejedor:gr-qc/0604031}
\bibinfo{author}{\bibfnamefont{A.}~\bibnamefont{{D{\'{\i}}ez-Tejedor}}}
  \bibnamefont{and}
  \bibinfo{author}{\bibfnamefont{A.}~\bibnamefont{{Feinstein}}},
  \bibinfo{journal}{\prd} \textbf{\bibinfo{volume}{74}},
  \bibinfo{pages}{023530} (\bibinfo{year}{2006}), \eprint{arXiv:gr-qc/0604031}.

\bibitem[{\citenamefont{{Chimento} and
  {Feinstein}}(2004)}]{Chimento:astro-ph/0305007}
\bibinfo{author}{\bibfnamefont{L.~P.} \bibnamefont{{Chimento}}}
  \bibnamefont{and}
  \bibinfo{author}{\bibfnamefont{A.}~\bibnamefont{{Feinstein}}},
  \bibinfo{journal}{Modern Physics Letters A} \textbf{\bibinfo{volume}{19}},
  \bibinfo{pages}{761} (\bibinfo{year}{2004}), \eprint{arXiv:astro-ph/0305007}.

\bibitem[{\citenamefont{{Hu} et~al.}(2000)\citenamefont{{Hu}, {Barkana}, and
  {Gruzinov}}}]{Hu:astro-ph/0003365}
\bibinfo{author}{\bibfnamefont{W.}~\bibnamefont{{Hu}}},
  \bibinfo{author}{\bibfnamefont{R.}~\bibnamefont{{Barkana}}},
  \bibnamefont{and}
  \bibinfo{author}{\bibfnamefont{A.}~\bibnamefont{{Gruzinov}}},
  \bibinfo{journal}{Physical Review Letters} \textbf{\bibinfo{volume}{85}},
  \bibinfo{pages}{1158} (\bibinfo{year}{2000}),
  \eprint{arXiv:astro-ph/0003365}.

\end{thebibliography}
\end{document}